%% file: main.tex
\documentclass[format=acmsmall, review=false, screen=true]{acmart} 

\usepackage{pifont}
\usepackage[most]{tcolorbox}
\usepackage[normalem]{ulem}
\usepackage{fontawesome}
\usepackage{threeparttable}
\usepackage{enumitem}
\usepackage{csquotes}
\usepackage{multirow}
\usepackage{url}
\usepackage{hyperref}

\newcommand{\ours}[1]{i\textsc{ReDev}}

\def\eg{\textit{e.g.,} }
\def\ie{\textit{i.e.,} }

\AtBeginDocument{%
  }

\setcopyright{acmlicensed}
\copyrightyear{2025}
\acmYear{2025}
\acmDOI{}
\acmJournal{TOSEM}
\acmISBN{}

\begin{document}





\title{\ours{}: A Knowledge-Driven Multi-Agent Framework for \underline{I}ntelligent \underline{Re}quirements \underline{Dev}elopment}


\author{Dongming Jin}
\email{dmjin@stu.pku.edu.cn}
\affiliation{%
  \institution{Peking University}
  \city{Beijing}
  \country{China}
}

\author{Weisong Sun}
\email{weisong.sun@ntu.edu.sg}
\authornote{Corresponding author}
\affiliation{
  \institution{Nanyang Technological University}
  \city{Singapore}
  \country{Singapore}
}

\author{Jiangping Huang}
\email{}
\affiliation{%
  \institution{Chongqing University of Posts and Telecommunications}
  \city{Chongqing}
  \country{China}
}

\author{Peng Liang}
\email{liangp@whu.edu.cn}
\affiliation{
  \institution{Wuhan University}
  \city{Wuhan}
  \country{China}
}

\author{Jifeng Xuan}
\email{jxuan@whu.edu.cn}
\affiliation{
  \institution{Wuhan University}
  \city{Wuhan}
  \country{China}
}

\author{Yang Liu}
\email{yangliu@ntu.edu.sg}
\affiliation{
  \institution{Nanyang Technological University}
  \city{Singapore}
  \country{Singapore}
}

\author{Zhi Jin}
\email{zhijin@pku.edu.cn}
\authornotemark[1]
\affiliation{
  \institution{Peking University and Wuhan University}
  \city{Beijing}
  \country{China}
}

\renewcommand{\shortauthors}{Jin et al.}

\input{chapters/0-abstract.tex}




\received{20 February 2007}
\received[revised]{12 March 2009}
\received[accepted]{5 June 2009}

\maketitle

\input{chapters/1-introduction.tex}

\input{chapters/2-background.tex}
\input{chapters/3-agent.tex}

\input{chapters/4-framework.tex}

\input{chapters/6-studydesign.tex}

\input{chapters/7-results.tex}

\input{chapters/9-discussion.tex}
\input{chapters/10-conclusion}

\bibliographystyle{ACM-Reference-Format}
\bibliography{main}

\end{document}

%% file: chapters/0-abstract.tex
\begin{abstract}
Requirements development is a critical phase in the software development life cycle as it is responsible for providing a clear understanding of what the end-users and stakeholders need. Requirement development goes beyond simply collecting information, but involves collaboration and communication, and critical thinking among stakeholders to extract explicit requirements, uncover hidden requirements, and address potential conflicts early in the project life cycle. This process is time-consuming and labor-intensive, and prone to errors.
With the emergence of large language models (LLMs), exploring LLM-based multi-agent systems for software development has attracted much attention. However, existing research provides limited support for requirements development and overlooks the injection of essential human knowledge into agent design and the critical role of human-agent collaboration. 

To address these issues, this paper proposes a knowledge-driven multi-agent framework for \underline{i}ntelligent \underline{re}quirement \underline{dev}elopment, named \ours{}. Unlike existing multi-agent frameworks, our framework features: \ding{182} \ours{} consists of six knowledge-driven agents (\ie interviewer, end-user, deployer, analyst, archivist and reviewer) to support the entire requirements development. They collaboratively perform various requirements development tasks (\ie elicitation, analysis, specification, and validation) to produce a well-defined software requirements specification. \ding{183} \ours{} specifically focuses on integrating the necessary human knowledge for agents, enabling them to simulate real-world stakeholders or requirements engineers to complete tedious requirements development tasks. \ding{184} \ours{} uses an event-driven communication mechanism based on a shared artifact pool that stores intermediate and final artifacts. Agents in \ours{} continuously monitor the artifact pool and autonomously trigger the next action based on its changes, enabling \ours{} to quickly handle new requirements and changes that may occur during the requirements development phase. \ding{185} \ours{} introduces a robust human-in-the-loop mechanism to support human-agent collaboration, ensuring that the generated artifacts align with the expectations of stakeholders. We perform experiments to evaluated the generated requirements artifacts (\eg requirements model and SRS) based on multiple traditional metrics and an LLM-as-a-judge-based metric. The results show that \ours{} outperforms existing baselines in multiple aspects. Following this framework, we further envision three key directions and hope this work can facilitate the development of intelligent requirements development.

\end{abstract}

\keywords{Requirements Development, Large Language Models, Multi-Agent Collaboration}

%% file: chapters/1-introduction.tex
\section{Introduction} \label{sec:01_introduction}

\begin{displayquote}
\textsf{Requirements development will become the most important work in software engineering when software reuse and automated programming make the expected progress.}  
\begin{flushright}
\footnotesize ------ Axel van Lamsweerde, \textit{ICSE 2000}~\citep{van2000requirements}; Douglas T Ross, \textit{TSE 1977}~\cite{ross1977structured}.
\end{flushright}
\end{displayquote}

The success of a software system depends on whether it can meet the needs of stakeholders and the constraints of the environment~\cite{cheng2007research}. 
Requirements development plays a crucial role in the software development life cycle, directly influencing the quality of a software system and the satisfaction of stakeholders~\cite{carroll1998requirements}. 
The purpose of requirements development is to derive and determine the appropriate and achievable requirements of stakeholders and the environment, which involves the activities of requirements elicitation~\cite{lim2011stakerare}, analysis~\cite{chechik2001automatic}, validation~\cite{ezzini2022automated}, and specification~\cite{jackson1995deriving}. These activities require experienced requirements engineers and close intervention from human stakeholders, making them time-consuming, labor-intensive, and prone to human biases and errors, particularly for large software projects~\cite{nuseibeh2000requirements}. Therefore, automating these activities is essential to improve software development productivity and reduce the burden on requirements engineers and human stakeholders. Meanwhile, with the rapid development of automated code generation and test generation, automated requirements development has become a critical bottleneck to further improve software development productivity~\cite{van2000requirements, ross1977structured}. 

Recently, large language models (LLMs) have achieved remarkable success across various individual software engineering tasks~\cite{hou2024large}, ranging from requirements development~\cite{khan2025large} and software design~\cite{white2024chatgpt,2025-MAAD} to code generation~\cite{jiang2024survey} and test case generation~\cite{wang2024software}. Simultaneously, LLMs demonstrate significant potential to achieve human-like intelligence~\cite{brown2020language}. Building upon this capability, LLM-based multi-agent systems offer opportunities to replicate human workflow and perform the entire software development process. 
For example, ChatDev~\cite{qian2023chatdev} structures the software development process into three phases (\ie designing, coding, and testing) and employs multiple specialized agents (\eg CEO and CTO) to contribute to these phases through native dialogue communication. It does not consider the requirements development phase, and its dialogue-based collaboration is complicated due to cascading hallucinations caused by natively chained LLMs~\cite{hong2024metagpt}. MetaGPT~\cite{hong2024metagpt} introduces a meta-programming framework for multi-agent collaboration and builds an agent-based software company to perform the entire software development process with the top-down waterfall model. AgileGen~\cite{zhang2024empowering} integrates agile methodologies for multi-agent systems to empower generative software development. Although MetaGPT and AgileGen involve the requirements development process, they simplify it into a requirements document generation task and lack systematic support for requirements elicitation, analysis, and validation. Elicitron~\cite{ataei2024elicitron} leverages multiple agents to automate the requirements elicitation, but only focuses on a single activity. MARE~\cite{jin2024mare} introduces a promising multi-agent framework to automate the entire requirements development process with LLMs. It emphasizes predefined profiles and actions of various agents for requirements development, lacks integration of human expert knowledge into agents, and overlooks the necessary human-agent collaboration.  

In summary, current multi-agent collaboration systems for software development provide limited support for requirements development due to the following four issues. 
\textbf{(1) The importance of requirements development is overlooked.} They either exclude the requirements development process or treat it simply as requirements document generation. In fact, requirements development is crucial to software development, which is a complex process involving multiple activities and the collaboration of multiple roles. 
\textbf{(2) The prior knowledge of human experts is ignored in agent design.} Existing works only assign agents as roles, without considering that various roles need to have the necessary expert knowledge. Injecting expert knowledge is a necessary condition for agents to act as various roles to perform complex requirements development activities, and this knowledge can provide a methodology and reasoning basis for agents' judgment. 
\textbf{(3) Current collaboration mechanisms struggle to cope with the requirements development process.} Current collaboration mechanisms in LLM-based agents do not align with the dynamic and interactive characteristics of requirements development. Specifically, mechanisms based on dialogue or the waterfall model are rigid and linear and cannot capture the essence of evolving requirements without feedback loops. 
\textbf{(4) The necessary intervention of human stakeholders is lacking.} 
Requirements development involves collecting the needs of real-world stakeholders. The participation of human stakeholders is crucial to provide key information and confirm that the requirements gathered align with the expectations of real-world stakeholders. The scope of requirements elicitation must be within the acceptable range of the customer's limited costs, such as time and money.

To address these issues, we propose a knowledge-driven multi-agent framework for \textbf{i}ntelligent \textbf{re}quirements \textbf{dev}elopment, called \ours{}.
First, \ours{} devises six agents to simulate real stakeholders involved in the requirements development phase, 
\ie interviewer agent, end user agent, deployer agent, analyst agent, archivist agent, and reviewer agent, specifically. Each agent is responsible for one or multiple requirements development activities and is equipped with the predefined actions to complete the activities. 
Second, to address the issue of ignoring expert knowledge in current agent design, we propose a new agent setting, \ie knowledge-driven agent. Specifically, we extract requirements-related expert knowledge (\eg thinking process and typical methodologies) for various roles from three primary sources (\ie professional books and project cases) and inject them into agents to play their roles using the chain-of-thought technique~\cite{wei2022chain}. 
Third, to solve the weakness of the current collaboration mechanism, \ours{} introduces an event-driven communication mechanism based on a shared artifact pool inspired by the blackboard mechanism~\cite{craig1988blackboard}. The shared artifact pool is designed for agents to upload intermediate requirements artifacts they have generated and retrieve the artifacts they need. Agents continuously observe the state of the artifact pool and autonomously trigger the next actions they need to take based on the state change (\ie an artifact is added or updated). This mechanism 
enhances the autonomy of agent collaboration and supports quick updates of requirement artifacts based on feedback from any stage. In addition, a robust human-in-the-loop (HITL) mechanism is integrated into our \ours{} to improve the reliability of the requirements development process and ensure that the produced requirements artifacts align with human stakeholders' expectations. 
Finally, based on the above systematic design, after giving a rough software requirements, \ours{} can iteratively perform a series of requirements development activities and communicate with humans to confirm and refine artifacts, thereby achieving autonomous high-quality requirements development.

To validate the effectiveness of \ours{}, we conduct experiments to evaluate intermediate requirements artifacts (\ie user requirements list (URL) and requirements model) and the final requirements specifications on 10 software systems from a public requirements development dataset. We employ multiple metrics (Section~\ref{sec:metrics}) to evaluate the quality of the produced artifacts and compare them with baselines (Section~\ref{sec:baselines}). The results show that \ours{} can deliver more diverse and balanced user requirements lists, construct more precise requirements models (\ie use case diagrams), and produce more high-quality and well-structured requirements specifications. 

\ours{} has demonstrated the feasibility of knowledge-driven multi-agent collaboration. Following this framework, we further envision three key topics: \textbf{(1) Automated Requirements Knowledge Extraction:} explores automatically discovering and optimizing the domain and procedural knowledge for a given requirements task. \textbf{(2) Automated Requirements Agent Generation:} generates agent prompts based on extracted knowledge to complete the requirements tasks. \textbf{(3) Automated Requirements Knowledge Evolution:} explores automatically detecting outdated, conflicting, or missing knowledge. We hope this work can provide a roadmap to facilitate the development of intelligent requirements development in the future.

Our contributions are as follows.
\begin{itemize}
    \item This paper proposes a knowledge-driven multi-agent framework for intelligent requirements development, which tackles the challenge of lacking enough support for the requirements development process. 
    \item This paper introduces a knowledge-driven agent design setting and extracts a systematic overview of knowledge for various roles and activities in the requirements development process. 
    \item This paper proposes an event-trigger communication mechanism based on a shared artifact pool, which enables quick updates of various requirements artifacts. 
    \item This paper introduces a robust human-in-the-loop mechanism to ensure that generated artifacts align with human stakeholders' expectations. 
    \item The experimental results have validated the efficiency of \ours{} in producing various requirements artifacts on 10 software projects.
\end{itemize}

\textit{Article Organization.} In the rest of the paper, Section~\ref{sec:02_background} presents the background and related work. Section~\ref{sec:03_agent_design} introduces the knowledge-driven agent design and presents an overview of various knowledge for the requirements development process. Section~\ref{sec:04_KANRE} presents the proposed \ours{} framework.
Section~\ref{sec:06_study_design} and ~\ref{sec:07_result} show the study design and experimental results, respectively. 
Section~\ref{sec:09_discussion} shows a case study and discusses the future directions.
Section~\ref{sec:10_conclusion} concludes this paper.

%% file: chapters/2-background.tex
\section{Background and Related Work} \label{sec:02_background}

\subsection{LLMs for automated Requirements Development}
Recently, researchers have started exploring the capabilities of LLMs in the requirements development process. We elaborate on the applications of LLMs in the activities of requirements elicitation, analysis, specification, and validation. 

\textbf{Requirements Elicitation.} Requirements elicitation is to derive the requirements or core functionalities from the stakeholders' needs of the software system and the constraints of the operating environment of the system. Researchers have used LLMs in various aspects to automatically elicit requirements.
For example, Elicitron~\cite{ataei2024elicitron} generates a diverse set of simulated user agents, which engage in product experience scenarios and undergo an agent interview process to surface latent user requirements. Their results show that Elicitron can effectively identify latent requirements and outperform traditional empathic lead user interviews. Gorer et al.~\cite{gorer2023generating} employed ChatGPT and Bard to generate interview scripts for requirements engineering training using few-shot and chain-of-thought prompting. They compared the generated interview scripts with those produced by human experts and found that the LLM-generated scripts were more effective and efficient. White et al.~\cite{white2024chatgpt} designed customized prompts and used ChatGPT to elaborate system requirements and underpin missing requirements adequately. Zhang et al.~\cite{zhang2024experimenting} leveraged ChatGPT to generate use cases by actively engaging different stakeholders to eliminate incompleteness and vagueness in the initial software requirements. Arora et al.~\cite{arora2024advancing} used ChatGPT to efficiently elicit software system requirements by addressing frequently occurring elicitation challenges, such as domain analysis. 

\textbf{Requirements Analysis.} The purpose of requirements analysis is to ensure the quality of raw stakeholders' requirements gathered during the elicitation phase. Researchers have proposed various LLM-based approaches to automate requirements analysis activity. Existing work can be divided into two categories, \ie LLMs for requirements text analysis and LLMs for requirements model extraction. The first category focuses mainly on clarifying and categorizing the raw elicited requirements for better understandability. For example, Ren et al~\cite{ren2024combining} proposed a prompt-based approach to classify end-user interviews into requirements and features using LLMs. Wei et al~\cite{wei2024requirements} proposed a progressive prompt approach to refine and generate detailed functional requirements based on relevant predecessor documents (\ie project glossary, vision and scope, and use cases). Khan et al.~\cite{khan2025leveraging} explored the potential of ChatGPT to classify end-user feedback into positive, negative, and natural sentiments. Feng et al.~\cite{feng2024normative} proposed a prompt-based approach to analyze non-functional requirements to overcome conflicting goals, priorities, and responsibilities. The second category involves constructing various requirements models from collected raw requirements to help understand user requirements and environment constraints. Recently, researchers have also started to use the power of LLMs to construct requirements models by generating various modeling diagrams.
For example, Ferrari et al.~\cite{ferrari2024model} explored the capability of ChatGPT to generate UML sequence diagrams from natural language requirements. Chen et al.~\cite{chen2023use} presented an exploratory study on the use of GPT-4 to create goal models. Chen et al.~\cite{chen2023automated} evaluated the ability of ChatGPT and GPT-4 to generate textual domain models from requirements descriptions using various prompting techniques (\ie zero-shot, k-shot, and chain-of-thought). Jin et al.~\cite{jin2024evaluation} proposed a requirements modeling benchmark and evaluated the performance of seven advanced LLMs in modeling the requirements of cyber-physical systems. 

\textbf{Requirements Specification.} The goal of requirements specification is to write software requirements in a formal and organized way, ensuring the consistency, completeness, and clarity of the requirements. Researchers have utilized LLMs to possibly automate requirements specification by generating, clarifying, and refining raw requirements into more structured requirements. Arora et al.~\cite{arora2024advancing} used ChatGPT to transform unstructured raw requirements into structured requirements with specific templates (\ie EARS or user stories). Leong et al.~\cite{leong2023translating} utilized ChatGPT and a symbolic method to transform natural language requirements into formal Java Modeling Language requirements. Lutze et al.~\cite{lutze2024generating} evaluated the performance of various LLMs in generating requirements specifications from requirements documents for smart devices. They found that LLMs can generate specifications with high accuracy but struggle with requirements that contain ambiguity or inconsistency. Jin et al.~\cite{jin2024mare} developed a documenter agent to write requirements specifications for a software project.   

\textbf{Requirements Validation.} Requirements validation is used to ensure quality attributes for requirements specifications. Researchers have started using LLMs to improve and automate this activity. For example, Helmeczi et al.~\cite{helmeczi2023few} proposed a few-shot learning approach to validate requirements specification documents to detect conflicts. Fantechi et al.~\cite{fantechi2023inconsistency} applied prompt engineering techniques and used ChatGPT to identify inconsistencies between requirements. Lubos et al.~\cite{lubos2024leveraging} explored the potential of LLMs to ensure the quality of software requirements in accordance with the quality characteristics defined in the ISO 29148 standard. 

Thus, LLMs have been widely used in various activities during the requirements development process. However, current strategies for using LLMs are mainly limited to simple prompt techniques, \eg zero-shot and few-shot. Specifically, they usually input task descriptions into the LLMs and ask them to generate responses directly. This strategy heavily relies on the internal knowledge of LLMs and lacks a reasoning process, resulting in unsatisfactory performance. Compared to them, our \ours{} focuses on extracting the prior expert knowledge on performing each activity and incorporating it into the LLMs using the chain-of-thought technique. They can guide LLMs in performing each activity more effectively and improving the reliability of the generated answers.

\subsection{Chain-of-Thought Prompting}
Chain-of-Thought (CoT) prompting~\cite{wei2022chain} has been a widely adopted and pivotal technique to improve the reasoning capabilities of LLMs. CoT involves prompting LLMs with a structured format of <input, thoughts, output>, where ``thoughts'' represents coherent intermediate reasoning steps from the ``input'' to the final answer (\ie ``output'')~\cite{li2023symbolic}. This approach has been demonstrated to be effective in various domains, such as arithmetic reasoning~\cite{wei2022chain}, and commonsense reasoning~\cite{li2023symbolic}. Thus, CoT has great potential to enhance the performance of LLMs in automating various requirements development activities. 

However, current strategies for automated requirements activities using LLMs are zero-shot and few-shot prompting, which tend not to achieve satisfactory performance. This is because they tend to provide direct answers without rationales, but these activities require a complex reasoning process. In contrast, CoT prompting can break down complex requirements development activities into manageable intermediate steps, which can guide LLMs through a logical progression. Specifically, the ``thoughts'' part can be considered as prior expert knowledge from requirements engineers in the real world. For example, it can be the thinking process of requirements engineers, the typical methodologies for requirements activities, and the criteria to be followed. To apply CoT prompting for requirements development activities, this prior knowledge should be extracted from various sources and incorporated into LLMs. Therefore, this paper first determines the required knowledge for various activities in the requirements development process (Section ~\ref{subsec:knowledge_list}), which can better set the agent profile and guide its reasoning and judgment. 

%% file: chapters/3-agent.tex
\section{Knowledge-Driven Agent} \label{sec:03_agent_design}

\subsection{Agent Design}

\begin{figure}
    \centering
    \includegraphics[width=0.5\linewidth]{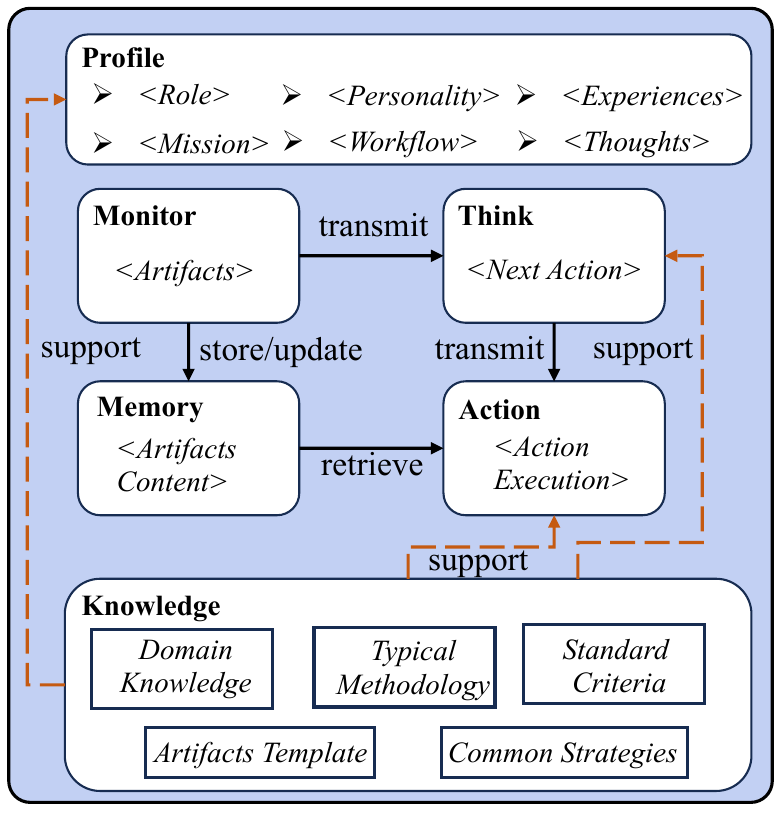}
    \caption{Overview of Knowledge-Driven Agent}
    \label{fig:knowledge_agent}
\end{figure}

The knowledge-driven agent is composed of five core modules, \ie profile, monitor, thinking, memory, action, and knowledge. As shown in Figure~\ref{fig:knowledge_agent}, these modules work together to enable the agent to interact effectively with various requirements artifacts and autonomously perform tasks related to requirements development. The details of these models are described in the following sections. 

\textbf{Profile.} The profile module defines the role and characteristics of the agent, allowing it to mimic real-world behaviors within the context of requirements development. The profile is written in the agent's system prompt and consists of three key parts, \ie personality information, experience information, and skill information. Specifically, the personality information includes the agent's role, personality, and mission. The experience information outlines the methodologies and workflow that the agent adheres to while performing tasks. The skill information details the thought process, strategies, and tools the agent uses to execute specific actions. 

\textbf{Monitor.} The monitor module is responsible for continuously overseeing the environment in which the agent operates. Specifically, it tracks changes (\ie additions and modifications) in requirements artifacts that the agent is concerned with. The monitors module allows the agent to stay aware of relevant shifts and notify the thinking module when state changes occur. The monitoring scope for each agent can be customized to include particular artifacts based on its role and goal, which ensures the agent focuses on the most pertinent requirements artifacts at any given time. 

\textbf{Thinking.} The thinking module aims to determine the next action to take dynamically based on the state change of requirements artifacts that the agent monitors. When the monitor module detects changes in the requirements artifacts, the thinking module judges whether an action is needed and determines which action should be taken. 
Specifically, the thinking module construct a prompt based on the prior knowledge from the knowledge module to guide the agent's reasoning, generating a series of intermediate steps to help the agent make the best decision. The thinking module enables the agent to make well-informed decisions by integrating its accumulated knowledge and the current state.

\textbf{Memory.} The memory module aims to store important information related to the requirements artifacts. The memory module allows the agent to retrieve this information quickly and facilitate subsequent processing by retaining the content and state of these artifacts. The memory module performs two key functions: memory writing and memory reading. The purpose of memory writing is to store information about the perceived artifacts in memory. 
The objective of memory reading is to extract stored meaningful information from memory to support the agent's action and decision-making process. This module ensures that the agent can work based on prior requirements artifacts, efficiently handling repetitive tasks without starting from scratch. 

\textbf{Action.} The action module is responsible for executing the tasks assigned to the agent. This module is located at the most downstream position and directly interacts with various requirements artifacts. A prompt should be carefully designed for each action, which define the specific task and thinking process to finish this task. Each action (\ie its prompt) is influenced by the thinking, memory, and knowledge modules. Actions can include processing raw requirements, generate requirements models, generating requirements specifications, and responding to stakeholders. The action module plays a key role in helping the agent achieve its goals.

\textbf{Knowledge.} The knowledge module stores the prior knowledge that the agent relies on when performing requirements development tasks, including methodologies to be followed and the thinking process involved in executing specific tasks. The module contains foundational knowledge, heuristic rules, best practices, and domain-specific principles, helping the agent execute the requirements development process. The knowledge module directly influences the profile, thinking, and action modules to provide rules, frameworks, and strategies for constructing their prompt, which ensures the agent can make appropriate responses when handling requirements development tasks. 

\subsection{Knowledge Extraction}~\label{subsec:knowledge_extraction}

The knowledge module is the core of the agent's expertise and decision-making capabilities. We outline the systematic extraction of requirements development knowledge from three primary sources: authoritative literature, existing requirements projects, and requirements experts. We detail the knowledge extraction process for each source. 

\textbf{Knowledge from Authoritative Literature.} Extracting knowledge from authoritative literature involves a collection and comprehensive analysis of textbooks, academic papers, and industry standards that focus on requirements development practices. They can provide typical methodologies and official guidelines to follow. 
The knowledge can be parsed and extracted manually or using natural language processing techniques. 
Specifically, it can include interview questions for requirements elicitation, definitions of meta-models for requirements modeling, standard templates for requirements specifications, and a checklist of requirements quality. Given that the extracted knowledge from this source is usually general, they are categorized and structured into a knowledge base that the agent can refer to during its tasks. The knowledge base enables the agent to apply proven methodologies and adhere to industry standards when performing requirements development tasks. 

\textbf{Knowledge from Existing Requirements Projects.} Extracting knowledge from existing requirements projects begins by gathering a comprehensive dataset of open-source or publicly available requirements development projects. They can provide insights about common development patterns, successful strategies, and valuable lessons learned from previous projects. The knowledge captured from these projects help to identify recurring issues, successful solutions, and effective practices proven in real-world scenarios. 
The knowledge can be extracted through manual analysis or text mining techniques. 
Considering that the extracted knowledge from this source can be continuously updated and refined as new projects are analyzed, they are structured into a dynamic knowledge base that the agent can refer to during its decision-making process. 

\textbf{Knowledge from Requirements Experts.} Extracting knowledge from requirements experts involves engaging with seasoned practitioners through interviews, surveys, and expert reviews to gather valuable insights and practical guidance. 
Structured knowledge elicitation techniques can be employed to gather and organize insights. 
During the process of knowledge elicitation, practitioners will be invited to share their experiences, methodologies, and insights on best practices based on their implicit knowledge. 
The knowledge from requirements experts can enhance the agent's adaptability, which allows it to handle complex and ambiguous scenarios that may not be covered by formal literature or existing projects.

\begin{table}[]
    \centering
    \setlength{\tabcolsep}{2pt}
    \caption{A Comprehensive Knowledge Map Across the Requirements Development Lifecycle}
    \begin{tabular}{ccccccc}
\toprule
\multirow{2}{*}{\textbf{Category}}   & \multirow{2}{*}{\textbf{Knowledge}} & \multirow{2}{*}{\textbf{Source}} & \multicolumn{4}{c}{\textbf{Requirements Development Process}}                           \\ \cmidrule{4-7} 
                                     &                                     &                                  & \textbf{Elicitation} & \textbf{Analysis} & \textbf{Specification} & \textbf{Validation} \\ \midrule
\multirow{3}{*}{Domain Knowledge}& Domain terminology / glossary & ~\cite{abad2018elica}& \faCheck & \faCheck & \faCheck & \faCheck\\
                                     & Industry processes / regulations & ~\cite{kenzi2010role}& \faCheck & \faCheck &  & \faCheck\\
 & NASA / FAA / V\&V cases & ~\cite{kapurch2010nasa}& \faCheck & \faCheck &  & \\ \midrule
\multirow{8}{*}{Typical Methodology}& Interviews / workshops  & ~\cite{Jin2023SRE_En}& \faCheck &  &  & \\
 & UML / SysML modeling & ~\cite{daoust2012uml}~\cite{infeld2018sysml}&  & \faCheck &  & \\
 & SysML / MBSE modeling & \href{https://github.com/Systems-Modeling/SysML-v2-Release}{Link 1} &  & \faCheck &  & \\
 & BPMN modeling & ~\cite{scholz2012business}&  & \faCheck &  & \\
 & Behavior‑driven specification & 
 \href{https://cucumber.io/}{Link 2}
 &  &  & \faCheck & \faCheck\\
 & Formal specification& ~\cite{merz2008specification}&  &  & \faCheck & \faCheck\\
 & Inspection/Peer Review)& ~\cite{firesmith2005quality}&  &  &  & \faCheck\\ 
 & Formal validation& ~\cite{cimatti2013validation}&  &  &  & \faCheck\\ \midrule
\multirow{5}{*}{Standards}& ISO/IEC/IEEE 29148 & 
~\cite{aware2018iso}
& \faCheck & \faCheck &  & \\
 & ISO/IEC 24744& 
 \href{https://en.wikipedia.org/wiki/ISO/IEC_24744}{Link 3}
 &  & \faCheck &  & \\
 & BPMN 2.0 & 
 \href{https://www.bpmn.org/}{Link 4}
 &  & \faCheck &  & \\
 & IEEE 1012‑2016& 
 ~\cite{ieee2016}
 &  &  &  & \faCheck\\
 & ISO 26262‑6& 
 ~\cite{iso26262}
 &  &  &  & \faCheck\\ \midrule
\multirow{7}{*}{Artifacts Template}& IEEE 830 SRS template & ~\cite{ieee830}&  &  & \faCheck & \faCheck\\
 & Use Case Specification& 
 \href{https://profinit.eu/wp-content/uploads/2016/03/use_case_template.doc}{Link 5}
 & \faCheck & \faCheck &  & \\
 & ReqIF‑based specification& 
 \href{https://www.omg.org/reqif/}{Link 6}
 &  &  & \faCheck & \faCheck\\
 & V\&V Plan Outline& 
 \href{https://www.nasa.gov/reference/appendix-i-verification-and-validation-plan-outline/}{Link 7}
 &  &  &  & \faCheck\\
 & Requirements Traceability Matrix& 
 ~\cite{Jin2023SRE_En}&  &  &  & \faCheck\\
 & Review checklists& ~\cite{Jin2023SRE_En}& & & &\faCheck\\ \midrule
\multirow{5}{*}{Common Strategies} & 5W1H & ~\cite{Jin2023SRE_En}&  \faCheck&  &  & \\
                                     & MoSCoW & \href{https://en.wikipedia.org/wiki/MoSCoW_method}{Link 8}&  &  \faCheck&  & \\
 & Socratic questioning & ~\cite{paul2019thinker}&  \faCheck&  &  & \\
 & Requirements Tradeoff& ~\cite{Jin2023SRE_En}&  &  &  \faCheck& \faCheck\\ \bottomrule
\end{tabular}
    \label{tab:knowledge_map}
\end{table}

\subsection{Knowledge List}~\label{subsec:knowledge_list}
 Table~\ref{tab:knowledge_map} demonstrates the extracted knowledge items we extracted and aligns them with the requirements development lifecycle, \ie elicitation, analysis, specification, and validation. We incorporate this knowledge for requirements agents in Section~\ref{sec:04_KANRE}. For clarity, the extracted knowledge is organized into five categories.

\begin{itemize}
    \item \textbf{Domain Knowledge:} defines the conceptual and regulatory context of the developed software system, which is required by the agent. It includes domain-specific terminology, glossary, industry processes, regulations, and safety-critical guidelines (\eg NASA, FAA and V\&V cases). By grounding decisions in this knowledge, the agent can recognize constraints such as real-time or certification requirements early, enabling more accurate analysis, specification, and validation.
    \item \textbf{Typical Methodologies:} offers procedural guidance across tasks from elicitation (\eg interviews, workshops) and modeling (\eg UML, SysML, and BPMN) to specification (\eg behavior-driven specification and formal languages) and validation (\eg peer review, inspection, and formal verification). This knowledge helps the agent adapt processes to project needs while ensuring methodological rigor.
    \item \textbf{Standards:} ensure outputs align with established best practices and regulatory norms. Key standards include ISO/IEC/IEEE 29148, ISO/IEC 24744, IEEE 1012-2016, and ISO 26262-6. The agent can check for completeness, traceability, and consistency, and offer corrective suggestions when deviations occur.
    \item \textbf{Artifact Templates:} provide reusable structures for efficient, consistent documentation. In addition to common templates (\eg IEEE 830 SRS), this includes specialized formats such as use case specification, ReqIF packages, traceability matrices, and peer review checklists. The agent selects and populates appropriate templates based on project context.
    \item \textbf{Common Strategies:} Equip the agent with universal reasoning techniques like 5W1H, MoSCoW prioritization, Socratic inquiry, and trade-off analysis. These strategies help uncover hidden assumptions, resolve conflicts, and align stakeholder interests. Plain-language transformation enhances clarity for non-technical stakeholders and supports cross-role communication.
\end{itemize}

%% file: chapters/4-framework.tex
\section{\ours{} Framework} \label{sec:04_KANRE}
In this section, we present a knowledge-driven multi-agent framework for intelligent requirements development, named \ours{}. We formally define the overview of our \ours{} and describe the detailing in the following sections, including three core modules.

\subsection{Overview}
\begin{figure*}[t]
    \centering
    \includegraphics[width=\linewidth]{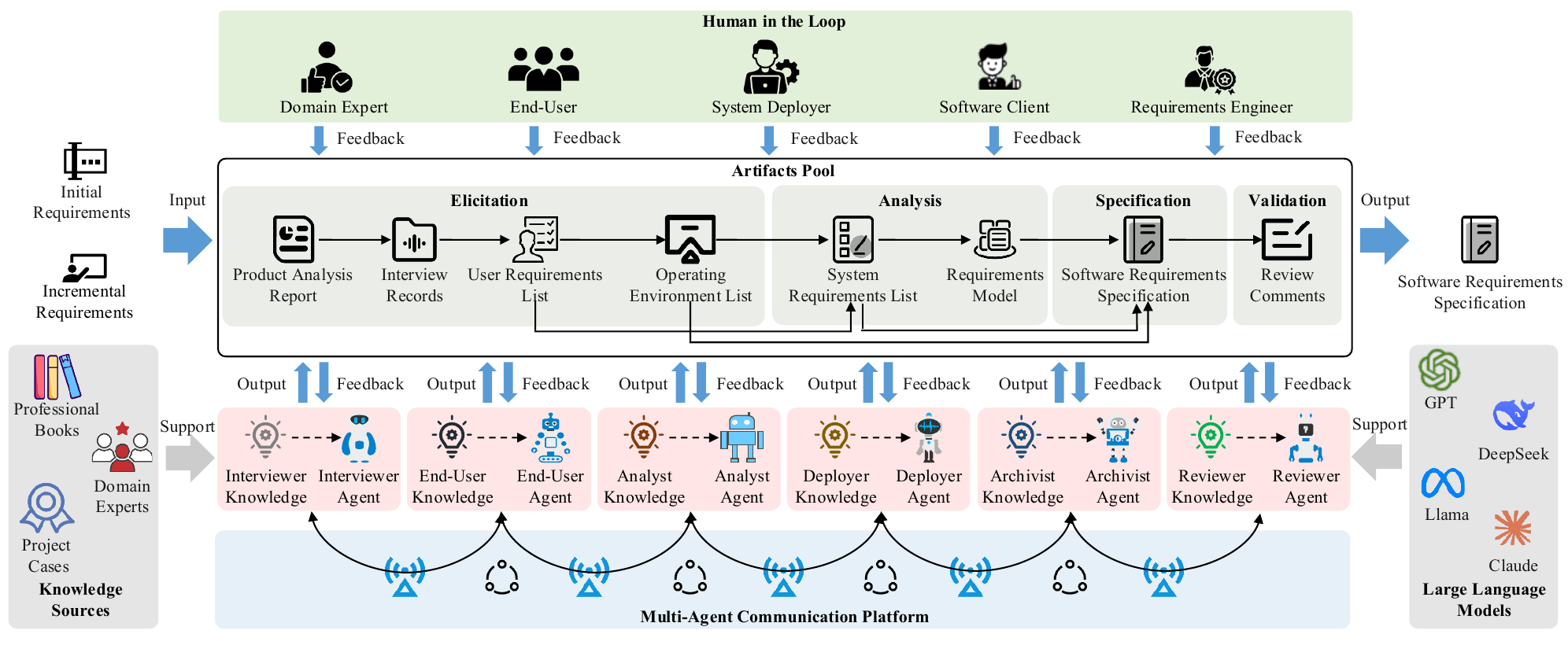}
    \caption{An overview of the knowledge-driven multi-agent framework for intelligent requirement development.}
    \label{fig:overview}
\end{figure*}

The goal of requirements development is to derive and generate an appropriate software requirements specification from stakeholders and the environment based on initial requirements for a new software system or incremental requirements for an existing software system. To achieve this goal, our \ours{} consists of three key modules as shown in Figure~\ref{fig:overview}, \ie six knowledge-driven agents, an artifacts pool, and a human-in-the-loop mechanism. 

\begin{itemize}
    \item \textbf{Six Knowldge-driven Agents.} These agents are designed to handle various tasks in the requirements development process autonomously. \ours{} includes six agents, \ie interviewer, enduser, deployer, analyst, archivist, and reviewer. The different designs for each agent are described in the following sections, including its profile, monitor, knowledge, and action. 
    \item \textbf{Artifact Pool.} The artifact pool serves as a central workspace to store all the intermediate and final artifacts generated during the requirements development process, which supports an event-trigger communication mechanism. It can ensure smooth communication and coordination between agents, allowing them to track changes in real-time and adjust their actions accordingly. 
    \item \textbf{Human-in-the-Loop.} The human-in-the-loop mechanism allows \ours{} to integrate feedback from various roles in the real world into the automated requirements development process. This ensures the generated requirements artifacts align with human stakeholders' expectations and enhances the overall quality of generated requirements artifacts. 
\end{itemize}

\subsection{Six Knowledge-driven Agents}
\ours{} includes six knowledge-driven agents, \ie interviewer, enduser, deployer, analyst, archivist, and reviewers. We describe the core modules for each agent. 

\begin{figure}[h]
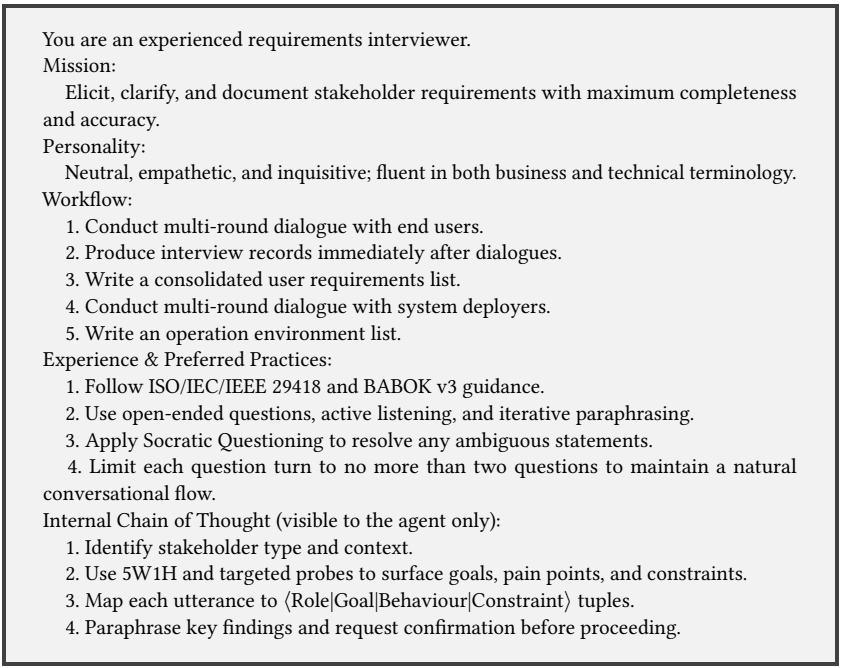

    \begin{tcolorbox}[sharp corners, width=0.8\textwidth]
    \footnotesize
    You are an experienced requirements interviewer. \\
Mission: \\
    \hspace*{0.8em} Elicit, clarify, and document stakeholder requirements with maximum completeness and accuracy. \\
Personality: \\
    \hspace*{0.8em} Neutral, empathetic, and inquisitive; fluent in both business and technical terminology. \\
Workflow: \\
    \hspace*{0.8em} 1. Conduct multi-round dialogue with end users. \\
    \hspace*{0.8em} 2. Produce interview records immediately after dialogues.\\
    \hspace*{0.8em} 3. Write a consolidated user requirements list.\\
    \hspace*{0.8em} 4. Conduct multi-round dialogue with system deployers.\\
    \hspace*{0.8em} 5. Write an operation environment list. \\
Experience \& Preferred Practices:\\
    \hspace*{0.8em} 1. Follow ISO/IEC/IEEE 29418 and BABOK v3 guidance.\\
    \hspace*{0.8em} 2. Use open-ended questions, active listening,  and iterative paraphrasing.\\
    \hspace*{0.8em} 3. Apply Socratic Questioning to resolve any ambiguous statements.\\
    \hspace*{0.8em} 4. Limit each question turn to no more than two questions to maintain a natural conversational flow. \\
Internal Chain of  Thought (visible to the agent only):\\
    \hspace*{0.8em} 1. Identify stakeholder type and context. \\
    \hspace*{0.8em} 2. Use 5W1H and targeted probes to surface goals, pain points, and constraints. \\
    \hspace*{0.8em} 3. Map each utterance to ⟨Role|Goal|Behaviour|Constraint⟩ tuples.\\
    \hspace*{0.8em} 4. Paraphrase key findings and request confirmation before proceeding. 
    \end{tcolorbox}
    \caption{Profile Prompt of the Interviewer agent}
    \label{fig:profile_prompt}
\end{figure}

\textbf{Interviewer Agent} aims to systematically elicit, clarify, and document user-level requirements from all relevant stakeholders (\eg enduser and deployer). The agent is prompted to follow the best practices in requirements elicitation, integrate domain-specific elicitation strategies, and international standards to ensure comprehensive and precise requirements capture. Figure~\ref{fig:profile_prompt} demonstrates the carefully constructed prompt for the interviewer agent's system prompt. \textbf{\textit{(1) Profile Design.}} The interviewer agent adopts a neutral, empathetic, and inquisitive tone to foster trust and openness during interactions with relevant stakeholders. Its mission is clearly framed around maximizing the completeness and accuracy of the elicited requirements. Its personality is tuned to navigate both technical and non-technical contexts with fluency. The workflow is explicitly structured into discrete steps: from engaging in dialogues with end users and deployers, to producing detailed interviewer records, consolidated requirements lists, and an operation environment list. The interviewer agent is prompted to follow ISO/IEC/IEEE 29148 and BABOK v3 standards and use some well-established techniques (\eg open-ended questioning, iterative paraphrasing, and socratic inquiry) into the agent. \textbf{\textit{(2) Monitor Design.}} The monitor module of the interviewer agent continuously observes upstream requirements artifacts, including the initial requirements description, interviewer records, and user requirements list. 
\textbf{\textit{(3) Knowledge Injection.}} 
The interviewer agent weaves together five complementary knowledge sources: industry terminology and regulations to surface implicit needs and compliance constraints; elicitation techniques (\eg 5W1H and Socratic questioning strategies) to structure its inquiries; international standards such as ISO/IEC/IEEE 29148 and BABOK v3 to ensure completeness and traceability; standardized document templates to keep records consistent; and dialogue strategies like MoSCoW prioritization and life-cycle trade-off reasoning to reconcile stakeholder priorities.
\textbf{\textit{(4) Predefined Actions.}} \ding{182} Dialogue with EndUser: generates the next question for the EndUser agent using the current dialogue context to surface goals, pain points, and constraints. \ding{183} Write Interview Records: consolidates the complete dialogue into a structured elicitation record. \ding{184} Write User‑Requirements List: synthesizes a hierarchical, prioritized list of user requirements, ensuring traceability to interview statements. \ding{185} Dialogue with Deployer: formulates targeted questions for the Deployer agent that probe infrastructure constraints, security mandates, and scalability expectations. \ding{186} Write Operating‑Environment List: compiles a comprehensive environment specification, capturing hardware, network, and compliance prerequisites derived from the deployer dialogue.  

\textbf{EndUser Agent} aims to simulate real end users, providing pain points, expectations, and usage feedback from a business scenario perspective. 
\textbf{\textit{(1) Profile Design.}} The EndUser agent is defined by a concise role description (e.g., sales clerk, warehouse supervisor), a set of daily tasks, and characteristic pain points. The agent adopts an approachable, conversational tone, emphasising goals and constraints rather than implementation detail. Emotional cues such as urgency or frustration are injected when relevant, mirroring real end‑user discourse. Although personas differ across domains, they all share a focus on business scenarios rather than system internals. \textbf{\textit{(2) Monitor Design.}} The EndUser agent remains dormant until it detects a new question from the interviewer agent.  
\textbf{\textit{(3) Knowledge Injection.}} The agent maintains a knowledge base organised into three layers. The scenario layer contains task workflows, business rules, and domain vocabulary that shape how goals and constraints are articulated. The pain‑point layer stores recurrent frustrations (\eg slow response times), enabling the agent to inject concrete cases. The quality‑expectation layer enumerates non‑functional concerns, \eg performance and data privacy requirements.
\textbf{(4) Predefined Actions.} \ding{182} Respond: provides goals, pain points, illustrative scenarios, and constraints that directly address the interviewer’s questions.
\ding{183} Raise Question: asks for clarification when the interview's question is ambiguous or conflicts with prior statements.
\ding{184} Confirm or Refine: validate earlier inputs or adjust them in light of new information, thereby supporting incremental, traceable consolidation of user‑level requirements.

\textbf{Deployer Agent} aims to articulate the technical and organisational constraints that shape how a system must be installed, configured, and maintained. 
\textbf{\textit{(1) Profile Design.}} Each Deployer persona is characterised by a specific hosting context (\eg database, network), and a strong focus on security and compliance. The agent communicates in a concise, technically focused style, foregrounding resource limits, network topology, access‑control policies, and automation pipelines. While pragmatic and risk‑averse, it remains cooperative, supplying concrete data that can be translated directly into an operating‑environment list. \textbf{\textit{(2) Monitor Design.}} The monitor module listens exclusively for targeted questions issued by the interviewer agent. 
\textbf{\textit{(3) Knowledge Injection.}} The Deployer agent is injected with environment-level elicitation knowledge, \ie domain terminology to articulate infrastructure components, ISO/IEC/IEEE 29148 standard that provides deploy environment checklist to ensure systematic coverage of configuration items, and requirements trade‑off strategies for balancing availability, cost, and performance. 
\textbf{\textit{(4) Predefined Actions.}} \ding{182} Respond: provides infrastructure constraints, security mandates, scalability targets, and operational procedures that answer the interviewer’s question.
\ding{183} Raise Question: requests clarification if a query lacks sufficient context.
\ding{184} Confirm or Refine: validate earlier deployment requirements or adjust them in light of new information.

\textbf{Analyst Agent} serves as the intellectual bridge between informal stakeholder statements and system requirements. It aims to distil user-centered and deployment-centered artifacts into a system requirements list and an appropriate requirements model. 
\textbf{\textit{(1) Profile Design.}} The agent projects a methodical, evidence‑based persona. It is familiar with both business and technical idioms and can converse with stakeholders while simultaneously reasoning in formal modeling terms. Its mission is to maximize requirement consistency, adhering to ISO/IEC/IEEE 29148 guidance for quality attributes and to IEEE 830 for specification. \textbf{\textit{(2) Monitor Design.}} The monitor module observes two upstream artifacts, \ie the user requirements list and operating environment list.  
\textbf{\textit{(3) Knowledge Injection.}} The agent leverages system requirements and requirements models knowledge, \ie IEEE 830 requirements template for drafting the system requirements list in a uniform style, UML/Sysml modeling meta-model to provide requirements modeling methodology.
\textbf{\textit{(4) Predefined Actions.}} \ding{182} Write System Requirements List: transforms the user‑level and environment-level requirements into a consolidated system requirements list.
\ding{183} Select Requirement Model: evaluates software system context to choose an appropriate modeling methodology (\eg use case diagram, and SysML diagram).
\ding{184} Build Requirement Model: construct requirements models with textual notation, highlighting conflicts or gaps for subsequent validation.

\textbf{Archivist Agent} functions as the project’s documentation curator, ensuring that every requirement is preserved in a cohesive software requirements specification. 
\textbf{\textit{(1) Profile Design.}} Projecting a meticulous and methodical persona, the Archivist speaks in a neutral, documentary tone.  Its core values are accuracy, completeness, and auditability.  
The agent also enforces naming conventions and metadata standards with IEEE 830 guidelines. \textbf{\textit{(2) Monitor Design.}} The monitor module observes two prerequisite artifacts: the system requirements list and the corresponding requirement Model. 
\textbf{\textit{(3) Knowledge Injection.}} To execute its archival duties, the agent draws on the IEEE 830 SRS template knowledge and ISO/IEC/IEEE 29148 documentation standard for structure, section headers, and required metadata. 
\textbf{\textit{(4) Predefined Actions.}} 
\ding{182} Write Software Requirements Specification: consolidates the approved system requirements list and requirement model into a standard structured SRS. 

\textbf{Reviewer Agent} acts as the project’s quality gatekeeper, ensuring that the software requirements specification satisfies accepted requirement quality criteria before it advances to design and implementation.  
\textbf{\textit{(1) Profile Design.}} It communicates with a formal, evidence-oriented tone, highlighting issues while offering actionable remediation advice. Its goal is not simply to approve or reject but to elevate the specification’s fitness for purpose, thereby reducing downstream rework and project risk. \textbf{\textit{(2) Monitor Design.}}  The monitor module observes the latest version of the SRS published by the archivist agent. 
\textbf{\textit{(3) Knowledge Injection.}} To achieve its duties, the agent leverages the requirements quality validation knowledge, \ie ISO/IEC/IEEE 29148 quality attributes (clarity, feasibility, verifiability, traceability, consistency), a peer‑review checklist template to ensure systematic coverage of each SRS section; and a catalogue of common requirement defects (ambiguity, conflict, redundancy). 
\textbf{\textit{(4) Predefined Actions.}} \ding{182} \textbf{Evaluate}: applies the checklist and quality criteria to evaluate the SRS, recording findings that cite specific sections and violated attributes.
\ding{184} \textbf{Confirm Closure}: examines the SRS after revisions to verify that all findings are resolved. 

\subsection{Artifacts Pool} 
The artifacts pool adopts a blackboard‑style, event‑driven architecture that stores every intermediate and final work product, from initial requirements to the definitive SRS. Any write or update operation emits a meta‑event broadcast that activates other agents’ monitors. This loosely coupled sharing mechanism can alleviate hallucination cascades, support parallel processing, and enable rapid rollback, thereby facilitating continuous feedback and incremental iteration across the lifecycle. Additionally, the artifact pool acts as a central workspace. It facilitates both communication and coordination among agents, each of which continuously monitors the pool for relevant changes and dynamically plans and executes actions based on the current state and content of artifacts. The requirements artifacts in this pool have five properties, which are content, role, state, sent\_from, and send\_to. The content property represents the content of the requirements artifacts. The role property indicates which agent generates the requirements artifacts. The state property describes whether this artifact has been modified. The sent\_from and send\_to properties give the flow of requirements artifacts among agents.

\subsection{Human in the Loop}
\ours{} bridges the gap between fully automated pipelines and real-world requirements by weaving a human-in-the-loop mechanism into every critical artifact hand-off. Whenever the user requirement list, requirements models, or SRS are generated, \ours{} conduct a pause for human confirmation and feedback. The feedback is written back to the shared artifacts pool, enabling downstream agents to parse revisions, which forms a closed loop of ``machine generation-human adjudication-machine correction''. This strategy balances automation speed with alignment to business goals and catches cascading errors without exhausting human bandwidth. \ours{} engages two types of human stakeholder roles in the HITL process, \ie requirements engineers and clients. The requirements engineers review various requirements artifacts generated by all agents. Their expert feedback is crucial for transforming raw requirements into an implementable, high-quality SRS. The clients focus on business alignment. Specifically, they verify whether the user requirements list captures their expectations and confirm business-oriented sections of the SRS (\eg Purpose).

%% file: chapters/6-studydesign.tex
\section{Study Design} \label{sec:06_study_design}
To assess the effectiveness of our \ours{}, we conduct an extensive study to answer three research questions. In this section, we describe the details of our study, including research questions, evaluation systems, metrics, and baselines. 

\subsection{Research Questions}
We illustrate the effectiveness of our \ours{} by answering the following three questions (RQs). 

\textbf{RQ1: How effective is \ours{} in producing the user requirements list during requirements development?} The user requirements list is a fundamental artifact generated after the elicitation phase, encapsulating the needs and expectations of stakeholders in a structured format. Investigating this RQ allows us to assess whether \ours{} can accurately extract and represent user needs based on initial, ambiguous, natural language descriptions.

\textbf{RQ2: How effective is \ours{} in producing requirements models during requirements development?}  Requirements modeling bridges the gap between informal requirements and formal system specifications. It enables structured reasoning, validation, and traceability. This RQ investigates whether \ours{} can transform textual requirements into formal or semi-formal models (\eg use case models) that are consistent, complete, and suitable for analysis. Evaluating the generated models helps determine the ability of to support model-driven development.

\textbf{RQ3: How effective is \ours{} in producing requirements specifications during requirements development?} The software requirements specification is a comprehensive artifact that consolidates user needs, system behaviors, constraints, and environment descriptions into a formal document that guides implementation. This RQ examines whether iReDev can integrate information from multiple upstream artifacts to produce coherent and complete specifications. As the SRS plays a pivotal role in bridging stakeholders and developers, its quality directly impacts system correctness and satisfaction.

\subsection{Evaluated Systems} ~\label{subsec: evaluated_system}
Our evaluation is based on ten real-world software systems created by users in previous work~\cite{zhang2024empowering}. Table~\ref{tab:evaluated_cases} shows the names and initial requirements of these systems. However, these software systems do not include reference user requirements lists, requirements models, and software requirements specifications. To obtain ground truth for simple evaluation, the first author manually constructs these corresponding artifacts for each software system. 

\begin{table}[]
    \centering
    \caption{Ten reality projects created by end-users for evaluation.}
    \begin{tabular}{lll}
\toprule
\textbf{ID} & \textbf{System Name}  & \textbf{Initial Requirements Description}                                                                                                        \\ \midrule
1 & Bookkeeping Assistant & I need a bookkeeping assistant website.                                                                                                          \\
2 & Random Roll Call      & I need a web system with a random roll call function.                                                                                              \\
3 & Shopping Site         & I need a shopping website.                                                                                                                       \\
4 & Gomoku                & Please design a basic Gomoku game.                                                                                                               \\
5 & Draw Flowers          & Please design a website that can draw different types of flowers.                                                                                \\
6 & Weather Forecast      & \begin{tabular}[c]{@{}l@{}}I need a weather display interface, which can show the weather condition of\\ city and future weather report.\end{tabular}   \\
7 & Online Timer          & \begin{tabular}[c]{@{}l@{}}I need a simple interface where users can set the duration of the\\ timer and start or stop the timer.\end{tabular} \\
8 & Currency Converter    & I need a currency converter webpage.                                                                                                    \\
9 & Online Translator     & Please generate an online translator website.                                                                                                    \\
10 & Event Reminder        & I would love to have a website that has added event reminder function                                                                            \\ \bottomrule
\end{tabular}
    \label{tab:evaluated_cases}
\end{table}

\subsection{Evaluation Metrics} ~\label{sec:metrics}
We use multiple traditional metrics to evaluate the quality of various artifacts in the above three RQs. These metrics are described in the following sections. 

\textbf{Metrics for User Requirements Lists.} Following the previous work~\cite{ataei2024elicitron}, two traditional metrics are employed to evaluate the diversity of user requirements lists. For all two metrics, we first generate embeddings for each requirements item in the user requirements list. Then we compute the \textit{convex hull volume (CHV)} and \textit{mean distance to centroid (MDC)}. 

\textbf{Metrics for Requirements Models.} Following the previous works on automated requirements modeling~\cite{chen2023automated}~\cite{jin2024evaluation}, We use five commonly used metrics to evaluate the quality of the requirements model. Specifically, the requirements model can be regarded as a set of nodes (\ie entity) and edges (\ie relationship). Thus the \textit{precision}, \textit{recall}, and \textit{F1 score} can be employed to evaluate the requirements model. The requirements model can also be transformed into a textual domain description (\eg PlantUML). Thus, the \textit{BLEU} and \textit{BertScore} can be used to evaluate the requirements models. 

\textbf{Metrics for Requirements Specifications.} Following the previous works on specification generation~\cite{dhulshette2025hierarchical}, we use traditional metrics (\ie \textit{BLEU}), transformer-based metrics (\ie \textit{BertScore}) and the LLM-as-a-judge in G-Eval~\cite{liu2023g}. With G-Eval, we consider the following criteria: (1) \textit{Completeness}: the generated SRS should cover all requirements in the ground truth; (2) \textit{Correctness}: the generated SRS should not hallucinate; 
and (3) \textit{Cohesiveness}: the generated SRS should be cohesive. 

\subsection{Baselines} ~\label{sec:baselines}
We select recently proposed automated approaches for requirements elicitation, modeling, and specification. 

\textbf{Common Baselines for Requirements Elicitation, Modeling and Specification.} We select 2 recently proposed common approaches as baselines. \textit{(1) LLM + zero-shot}: uses a single LLM (\eg GPT-4) that receives only the problem statement and an instruction prompt that specifies that the expected output format (\ie user requirements list, UML requirements model, and SRS). No examples, role descriptions, or intermediate decomposition steps are provided. In our experiments, we apply the same domain‑independent template across all three artifacts. Only the <task‑type> prompt token is varied (\ie URL, Model, SRS) to signal the desired artifact output. \textit{(2) LLM + MetaGPT}: MetaGPT is a multi‑agent framework that encodes standardised operating procedures (SOPs) as a chain of role‑specific prompts (\eg Product Manager, Engineer, QA) and orchestrates them as an ``assembly line" to reduce cascading hallucinations during complex software engineering tasks. In our experiments, we use GPT-4 and apply its pretrained prompt for requirements development to generate various requirements artifacts.

\textbf{Additional Baselines for Requirements Elicitation.} We select 1 recently proposed multi-agent requirements elicitation approach (\ie Elicitron) as an additional baseline for requirements elicitation. Elicitron is a recent LLM‑driven framework that simulates a diverse population of virtual end‑users and conducts empathic interviews to surface both explicit and latent needs. We adopt the public implementation and run one interviewer agent against ten simulated users per scenario, following the authors’ recommended hyper‑parameters.  The resulting consolidated need statements are treated as the users requirements list for evaluation.

\subsection{Experiment Settings.} We use the advanced \textit{GPT-4-turbo-2024-04-09} as the base LLM of \ours{}. 
To ensure the stability of the generated artifacts, the output confidence parameter \textit{Top-P} is set to 1.0 using the default value, and the frequency and presence penalties are set to 0.0. The output randomness parameter \textit{Temperature} is set to 0.3, and the \textit{maximum token size} is set to 4096. Besides, we set the artifact pool empty. 

%% file: chapters/7-results.tex
\section{Results and Analyses} \label{sec:07_result}

\textbf{RQ1: How effective is \ours{} in producing the user requirements list during requirements development?} 

\textbf{Setup.} We first input the initial requirements descriptions into \ours{} and use it to generate user requirements lists for the selected systems. We evaluate our \ours{} and three baselines (Section~\ref{sec:baselines}) on 10 software systems (Table~\ref{tab:evaluated_cases}). The evaluation metrics are described in Section~\ref{sec:metrics}, \ie CHV and MDC. For all metrics, higher scores represent better performance. 

\textbf{Results.} Table~\ref{tab:rq1} demonstrates the experimental results on the quality of generated user requirements lists for ten software systems. 

\textbf{Analyse.} \textbf{(1) Overall superiority.} As shown in Table~\ref{tab:rq1}, \ours{} achieves the highest scores on both diversity metrics across all ten evaluated systems. Its average CHV reaches 0.47, surpassing the strongest baseline (GPT‑4 + Elicitron) by 46.9\%, and its average MDC climbs to 0.62 with an improvement of 12.7\%. These results indicate that \ours{} can produce user requirement lists that cover a broader semantic space and exhibit a more even distribution than existing approaches. \textbf{(2) Consistent robustness.} The advantage of \ours{} is consistent on every individual system from the smallest (System 3) to the most complex (System 7). Compared with the zero‑shot baseline, the average CHV rises by 262\% and the average MDC by 55\%. Compared with the multi-agent‑based approach MetaGPT, \ours{} still boosts CHV by 124\% and MDC by 24\%. This uniform superiority demonstrates that \ours{} adapts stably to projects of varying domains and scales.

\input{tables/rq1}

\input{tables/rq2}

\textbf{RQ2: How effective is \ours{} in producing requirements models during requirements development?}

\textbf{Setup.} We first use \ours{} to generated requirements models (Use Case Diagram with PlantUML) for selected systems. 
Then we evaluate two baselines (Section~\ref{sec:baselines}) and our \ours{} on 10 software systems (Table~\ref{tab:evaluated_cases}). The evaluation metrics are described in Section~\ref{sec:metrics}, \ie F1, BLUE, and BertScore. For all metrics, higher scores represent better performance. 

\textbf{Results.} Table~\ref{tab:rq2} shows the experimetnal results on the quality of generated requirements models (use case diagrams) for the ten software systems.

\textbf{Analyses.} Overall, \ours{} significantly outperforms both baselines in generating high-quality requirements models, as indicated by the average scores across all three metrics. Specifically, \ours{} achieves the highest average F1 score (0.389), which is over three times higher than that of GPT-4 + MetaGPT (0.109) and nearly an order of magnitude higher than GPT-4 + zero-shot (0.025). This indicates that \ours{} is more accurate in identifying and generating correct use case elements, such as actors and use cases, when compared to the ground truth. In terms of semantic fidelity, \ours{} achieves the best average BertScore (0.593), suggesting that its generated use case diagrams are more semantically similar to the ground truth representations. Compared to the BertScore of GPT-4 + MetaGPT (0.442) and GPT-4 + zero-shot (0.387), the improvements imply that \ours{} is better at capturing the meaning and structure of user requirements, rather than just surface-level tokens. Although BLEU scores are generally low across all approaches due to the structural sparsity and variability of PlantUML representations, \ours{} still achieves a substantial relative improvement (0.102) over GPT-4 + MetaGPT (0.063) and GPT-4 + zero-shot (0.045). This suggests that \ours{} generates more consistent and n-gram-overlapping content with the ground truth, even under token-level evaluation. These results collectively demonstrate that \ours{} is more effective in producing high-fidelity and structurally accurate requirements models.

\input{tables/rq3}

\textbf{RQ3: How effective is \ours{} in producing requirements specifications during requirements development?}

\textbf{Setup.} We also first use \ours{} to generated software requirements specifications on the same systems. Then we evaluate two baselines (Section~\ref{sec:baselines}) and our \ours{} on 10 software systems (Table~\ref{tab:evaluated_cases}). The evaluation metrics are described in Section~\ref{sec:metrics}, \ie BLUE, BertScore, and G-Eval based on three criteria (Completeness, Correctness, and Cohesiveness).

\textbf{Results.} Table \ref{tab:rq3} shows the results of evaluation for requirements specifications generated by \ours{}.

\textbf{Analyses.} Our \ours{} significantly outperforms the two baseline methods (zero-shot and MetaGPT) across all five evaluation metrics. Specifically, \ours{} achieves the highest average BLEU score (0.120), indicating better alignment with the reference specifications in terms of n-gram overlap. Likewise, the BertScore for \ours{} reaches 0.616 on average, surpassing both baselines and reflecting higher semantic similarity between the generated and reference texts. In terms of LLM-as-a-Judge evaluation based on the G-Eval, \ours{} consistently demonstrates superior performance. The average completeness score improves from 2.900 (zero-shot) and 3.200 (MetaGPT) to 4.200 with \ours{}, showing its enhanced capability in covering essential requirement elements. Similarly, the correctness score increases from 2.400 and 3.000 to 4.000, indicating a higher degree of factual accuracy in the generated content. The cohesiveness metric also shows a notable gain, rising from 3.200 and 3.700 to 4.100, suggesting that \ours{} produces more logically structured and coherent requirement documents.Overall, these results validate the effectiveness of \ours{} in generating high-quality software requirements specifications. 
The improvements across automatic and manual evaluation metrics affirm the benefits of integrating domain knowledge and development-phase context through the \ours{} framework.

%% file: tables/rq1.tex
\begin{table}[]
    \centering
    \caption{Results on the Diversity of User Requirements Lists}
    \begin{tabular}{ccccccccccccc}
\toprule
\multirow{2}{*}{\textbf{\begin{tabular}[c]{@{}c@{}}System\\ ID\end{tabular}}} 
      & \multicolumn{2}{c}{\textbf{GPT‑4 + zero‑shot}} 
      & \multicolumn{2}{c}{\textbf{GPT‑4 + MetaGPT}} 
      & \multicolumn{2}{c}{\textbf{GPT‑4 + Elicitron}} 
      & \multicolumn{2}{c}{\textbf{GPT‑4 + \ours{}}} \\ \cmidrule(lr){2-3}\cmidrule(lr){4-5}\cmidrule(lr){6-7}\cmidrule(lr){8-9}
      & CHV  & MDC  & CHV  & MDC  & CHV  & MDC  & CHV  & MDC  \\ \midrule
1  & 0.12 & 0.38 & 0.19 & 0.48 & 0.30 & 0.54 & 0.45 & 0.60 \\
2  & 0.14 & 0.40 & 0.20 & 0.50 & 0.32 & 0.55 & 0.47 & 0.62 \\
3  & 0.11 & 0.37 & 0.18 & 0.47 & 0.28 & 0.52 & 0.42 & 0.58 \\
4  & 0.15 & 0.42 & 0.22 & 0.52 & 0.35 & 0.56 & 0.50 & 0.64 \\
5  & 0.13 & 0.39 & 0.21 & 0.50 & 0.34 & 0.55 & 0.48 & 0.62 \\
6  & 0.12 & 0.38 & 0.20 & 0.49 & 0.31 & 0.54 & 0.46 & 0.61 \\
7  & 0.16 & 0.43 & 0.24 & 0.53 & 0.36 & 0.57 & 0.52 & 0.65 \\
8  & 0.14 & 0.41 & 0.22 & 0.51 & 0.33 & 0.56 & 0.49 & 0.63 \\
9  & 0.13 & 0.40 & 0.21 & 0.50 & 0.32 & 0.55 & 0.48 & 0.62 \\
10 & 0.12 & 0.39 & 0.19 & 0.48 & 0.30 & 0.53 & 0.45 & 0.60 \\ \midrule
\textbf{Ave} 
   & \textbf{0.13} & \textbf{0.40} 
   & \textbf{0.21} & \textbf{0.50} 
   & \textbf{0.32} & \textbf{0.55} 
   & \textbf{0.47} & \textbf{0.62} \\ \bottomrule
    \end{tabular}
    \label{tab:rq1}
\end{table}

%% file: tables/rq2.tex
\begin{table}[]
    \centering
    \caption{Results on the Quality of Requirements Models (Use‑Case Diagrams)}
    \begin{tabular}{cccccccccc}
\toprule
\multirow{2}{*}{\textbf{\begin{tabular}[c]{@{}c@{}}System\\ ID\end{tabular}}} 
        & \multicolumn{3}{c}{\textbf{GPT‑4 + zero‑shot}} 
        & \multicolumn{3}{c}{\textbf{GPT‑4 + MetaGPT}} 
        & \multicolumn{3}{c}{\textbf{GPT‑4 + \ours{}}} \\ \cmidrule(lr){2-4}\cmidrule(lr){5-7}\cmidrule(lr){8-10}
        & F1   & BLEU  & BertScore 
        & F1   & BLEU  & BertScore 
        & F1   & BLEU  & BertScore \\ \midrule
1   & 0.00 & 0.04 & 0.50 & 0.10 & 0.06 & 0.55 & 0.40 & 0.10 & 0.65 \\
2   & 0.05 & 0.03 & 0.44 & 0.12 & 0.05 & 0.47 & 0.35 & 0.09 & 0.60 \\
3   & 0.20 & 0.04 & 0.29 & 0.35 & 0.07 & 0.40 & 0.60 & 0.12 & 0.55 \\
4   & 0.00 & 0.05 & 0.42 & 0.08 & 0.07 & 0.49 & 0.38 & 0.11 & 0.62 \\
5   & 0.00 & 0.03 & 0.47 & 0.07 & 0.05 & 0.51 & 0.37 & 0.09 & 0.64 \\
6   & 0.00 & 0.04 & 0.44 & 0.09 & 0.06 & 0.49 & 0.36 & 0.10 & 0.63 \\
7   & 0.00 & 0.11 & 0.11 & 0.05 & 0.13 & 0.20 & 0.30 & 0.15 & 0.45 \\
8   & 0.00 & 0.04 & 0.43 & 0.08 & 0.05 & 0.46 & 0.40 & 0.09 & 0.61 \\
9   & 0.00 & 0.04 & 0.43 & 0.08 & 0.05 & 0.45 & 0.38 & 0.09 & 0.60 \\
10  & 0.00 & 0.03 & 0.34 & 0.07 & 0.04 & 0.40 & 0.35 & 0.08 & 0.58 \\ \midrule
\textbf{Ave} 
    & \textbf{0.025}  & \textbf{0.045}  & \textbf{0.387}  
    & \textbf{0.109}  & \textbf{0.063}  & \textbf{0.442}  
    & \textbf{0.389}  & \textbf{0.102}  & \textbf{0.593}  \\ \bottomrule
    \end{tabular}
    \label{tab:rq2}
\end{table}


%% file: tables/rq3.tex
\begin{table}[]
    \centering
    \caption{Results on the Quality of Requirements Specifications}
    \begin{tabular}{cccccccccccc}
\toprule
\multirow{2}{*}{\textbf{Metrics}} & \multicolumn{11}{c}{\textbf{System ID}}                                                                    \\ \cmidrule{2-12}
                                  & \textbf{1} & \textbf{2} & \textbf{3} & \textbf{4} & \textbf{5} & \textbf{6} & \textbf{7} & \textbf{8} & \textbf{9} & \textbf{10} & \textbf{Ave} \\ \midrule
\multicolumn{12}{c}{\textbf{GPT‑4 + zero‑shot}}                                                                                               \\ \midrule
BLEU                              & 0.06 & 0.06 & 0.04 & 0.09 & 0.06 & 0.06 & 0.10 & 0.08 & 0.04 & 0.04 & \textbf{0.063} \\
BertScore                         & 0.54 & 0.50 & 0.50 & 0.55 & 0.53 & 0.51 & 0.54 & 0.47 & 0.52 & 0.53 & \textbf{0.519} \\
Completeness                      & 3    & 2    & 3    & 3    & 3    & 3    & 3    & 3    & 3    & 3    & \textbf{2.900} \\
Correctness                       & 3    & 2    & 2    & 4    & 2    & 2    & 2    & 3    & 2    & 2    & \textbf{2.400} \\
Cohesiveness                      & 4    & 3    & 3    & 4    & 3    & 3    & 3    & 3    & 3    & 3    & \textbf{3.200} \\ \midrule
\multicolumn{12}{c}{\textbf{GPT‑4 + MetaGPT}}                                                                                                 \\ \midrule
BLEU                              & 0.08 & 0.07 & 0.06 & 0.10 & 0.08 & 0.08 & 0.11 & 0.09 & 0.07 & 0.06 & \textbf{0.080} \\
BertScore                         & 0.56 & 0.55 & 0.54 & 0.57 & 0.56 & 0.55 & 0.57 & 0.53 & 0.54 & 0.55 & \textbf{0.552} \\
Completeness                      & 3    & 3    & 3    & 4    & 3    & 3    & 3    & 4    & 3    & 3    & \textbf{3.200} \\
Correctness                       & 3    & 3    & 2    & 4    & 3    & 3    & 3    & 3    & 3    & 3    & \textbf{3.000} \\
Cohesiveness                      & 4    & 4    & 3    & 4    & 4    & 3    & 4    & 4    & 3    & 4    & \textbf{3.700} \\ \midrule
\multicolumn{12}{c}{\textbf{GPT‑4 + \ours{}}}                                                                                                 \\ \midrule
BLEU                              & 0.12 & 0.11 & 0.10 & 0.14 & 0.12 & 0.12 & 0.15 & 0.13 & 0.11 & 0.10 & \textbf{0.120} \\
BertScore                         & 0.62 & 0.61 & 0.60 & 0.63 & 0.62 & 0.61 & 0.64 & 0.60 & 0.61 & 0.62 & \textbf{0.616} \\
Completeness                      & 4    & 4    & 4    & 5    & 4    & 4    & 4    & 5    & 4    & 4    & \textbf{4.200} \\
Correctness                       & 4    & 4    & 3    & 5    & 4    & 4    & 4    & 4    & 4    & 4    & \textbf{4.000} \\
Cohesiveness                      & 4    & 4    & 4    & 5    & 4    & 4    & 4    & 4    & 4    & 4    & \textbf{4.100} \\ \bottomrule
    \end{tabular}
    \label{tab:rq3}
\end{table}


%% file: chapters/9-discussion.tex
\section{Discussion} \label{sec:09_discussion}


\begin{figure*}[!t]
    \centering
    \includegraphics[width=0.98\linewidth]{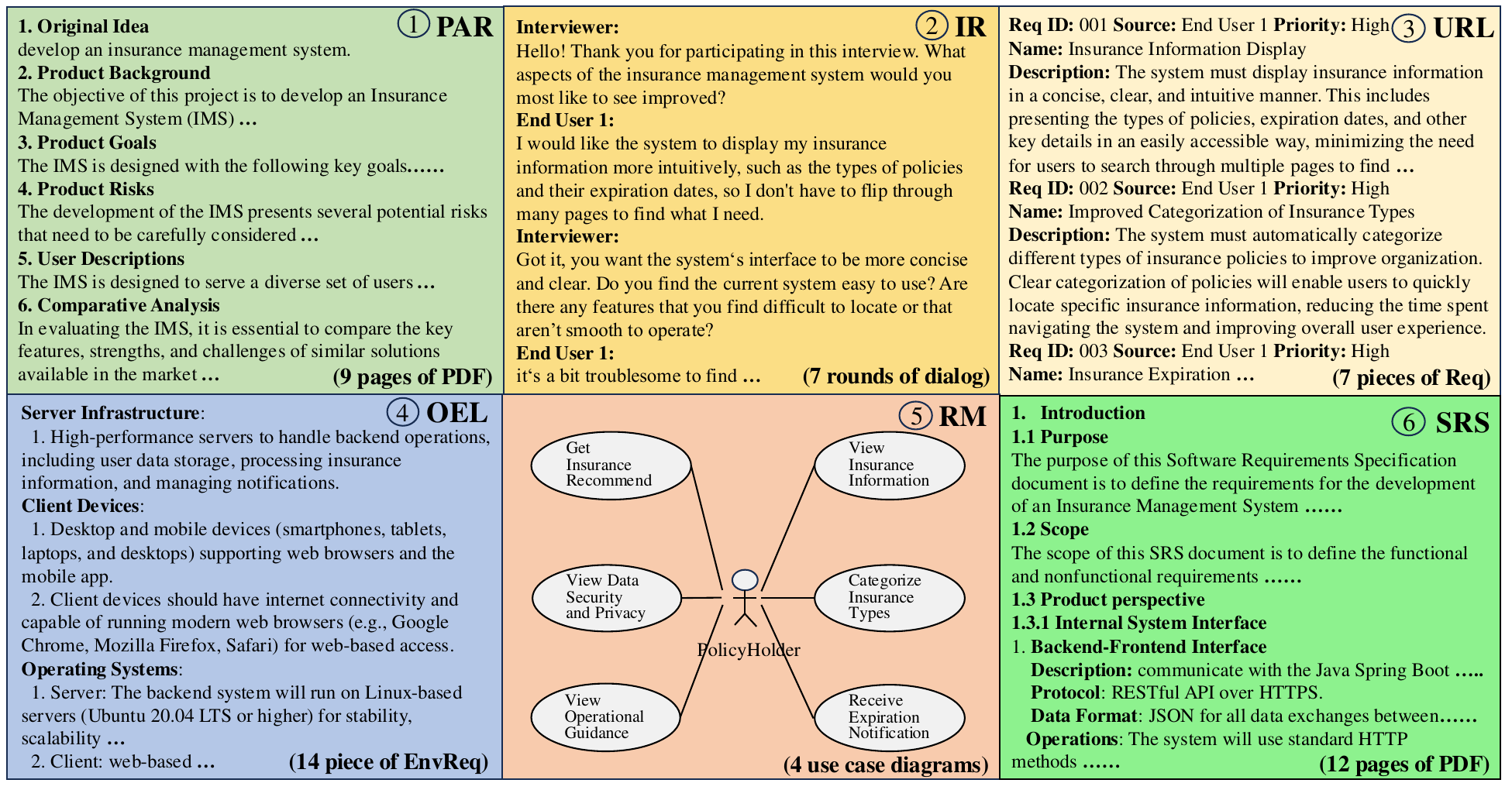}
    \vspace{-4mm}
    \caption{The artifacts generated by \ours{} on the insurance management system. 
    }
    \label{fig:case study}
    \vspace{-2mm}
\end{figure*}

\subsection{Case Study}
\textbf{The Selected Case.} The selected case is an enterprise-level web-based insurance management system. This system is designed to offer client information management services to insurance companies, primarily to distribute employee welfare insurance and to store related data. The reason for choosing this case is that it demonstrates a real-world business scenario with wide application. Requirements development for such an application involves various requirements-related tasks. 

\textbf{Generated Artifacts.} Figure~\ref{fig:case study} presents partial content of the six major artifacts generated in this case study. It can be observed that the quality of both the intermediate artifacts and the final SRS, produced through the collaboration of multiple agents within \ours{}, is satisfactory. Our framework supports HITL, but we report fully automated collaboration to establish the baseline. All results, including HITL variants, can be accessed via our public link~\cite{web:code}.

\subsection{Threat to validity}
\textbf{Construct Validity} concerns the relationship between treatment and outcome. The threat comes from the rationality of the research questions we asked. Our study operationalises the three research questions (RQ1–RQ3) through a collection of automatic metrics. The main threat is that these proxies may not fully capture the latent constructs we intend to measure. \textit{(1) Semantic coverage vs. stakeholder value}. CHV and MDC assume that a broader embedding space equates to a richer requirements set, but high diversity does not guarantee that truly relevant needs are captured. \textit{(2) Token‑level vs. structural fidelity}. BLEU and BertScore reward n‑gram or semantic overlap, but ignore diagram layout or trace links that practitioners care about in models and specifications. \textit{(3) LLM‑as‑Judge subjectivity}. G‑Eval inherits the biases of the underlying LLM and its prompt. Disagreement with human experts is possible. We mitigated these risks by triangulating multiple metrics drawn from prior RE literature

\textbf{Internal Validity} addresses potential threats to the way the study was conducted. This validity threat arises from choices in our experimental procedure. The threat comes from the golden artifact construction of our evaluated systems (Section~\ref{subsec: evaluated_system}). To obtain ground truth for evaluation, the first author manually constructs corresponding artifacts. In this process, we acknowledge that their annotations by hand are somewhat subjective. To mitigate this threat, we invited an external requirements engineering practitioner to double-review the gold artifacts.

\textbf{External Validity} considers the generalizability of our findings. The first threat is the project scale and domain of our selected evaluation systems. The ten target systems are small‑to‑medium web or desktop applications written in English. Industrial, safety‑critical or multilingual projects could expose additional challenges (e.g., domain jargon, regulatory constraints). The second threat is the selected LLMs for experiments. \ours{} currently focuses on GPT‑4‑turbo and an English knowledge corpus. Performance may vary with smaller LLMs, different languages or cross‑cultural stakeholders. 

\subsection{Future Directions}  
\ours{} has demonstrated the feasibility of knowledge-driven multi-agent collaboration for intelligent requirements development, significantly enhancing the quality of requirements artifacts. However, the current framework is still a prototype that (1) depends on manually extracting and injecting knowledge for each agent and task and (2) treats knowledge as a static asset. Thus, future research can focus on the following key topics:

\begin{itemize}
    \item \textbf{Automated Requirements Knowledge Extraction (From Task to Knowledge).} Given a requirements task, extracting the required knowledge by hand is time-consuming and labor-intensive. Thus, future work can explore automatically discovering and optimizing the domain and procedural knowledge needed for each task. For example, the knowledge can be parsed and extracted from authoritative literature using natural language processing techniques, and from existing projects through text mining techniques.
    \item \textbf{Automated Requirements Agent Generation (From Knowledge to Agent Prompt).} Once the required knowledge is identified, the next step is to generate agent prompts to complete the task. Future work can focus on designing automatic pipelines that translate structured knowledge (\eg domain rules, best practices, requirement templates) into optimized prompt templates. This process may involve knowledge-to-text generation, context-aware template selection, and multi-turn prompt planning. 
    \item \textbf{Automated Requirements Knowledge Evolution.} Requirement domains are inherently dynamic, \ie new knowledge emerges, and existing knowledge becomes outdated. Future research can explore mechanisms for detecting outdated, conflicting, or missing knowledge. This may involve continual learning or incremental update strategies to allow the knowledge base to evolve over time. 
\end{itemize}

%% file: chapters/10-conclusion.tex
\section{Conclusion}~\label{sec:10_conclusion}
This paper presents \ours{}, a knowledge‑driven multi‑agent framework with human‑in‑the‑loop support that unifies requirements elicitation, analysis, specification and validation. Experiments on ten real‑world software projects show that \ours{} consistently outperforms state‑of‑the‑art baselines in the quality of user requirement lists, use‑case models, and software requirements specifications, demonstrating both effectiveness and generality. Industrial case studies further confirm that the framework can produce large-scale and satisfactory artifacts. These results demonstrated the feasibility of knowledge-driven and collaboration mechanisms. We further envision three key topics following this framework. Additionally, we hope this framework can provide a roadmap to facilitate the development of intelligent requirements development in the future.